# A block EM algorithm for multivariate skew normal and skew $t$-mixture models

Sharon X. Lee[1], Kaleb L. Leemaqz[1], Geoffrey J. McLachlan[1,⋆]

[1]Department of Mathematics, University of Queensland, St. Lucia, Queensland, 4072, Australia.
⋆ E-mail: g.mclachlan@uq.edu.au

**Abstract**

Finite mixtures of skew distributions provide a flexible tool for modelling heterogeneous data with asymmetric distributional features. However, parameter estimation via the Expectation-Maximization (EM) algorithm can become very time-consuming due to the complicated expressions involved in the E-step that are numerically expensive to evaluate. A more time-efficient implementation of the EM algorithm was recently proposed which allows each component of the mixture model to be evaluated in parallel. In this paper, we develop a block implementation of the EM algorithm that facilitates the calculations in the E- and M-steps to be spread across a larger number of threads. We focus on the fitting of finite mixtures of multivariate skew normal and skew $t$-distributions, and show that both the E- and M-steps in the EM algorithm can be modified to allow the data to be split into blocks. The approach can be easily implemented for use by multicore and multiprocessor machines. It can also be applied concurrently with the recently proposed multithreaded EM algorithm to achieve further reduction in computation time. The improvement in time performance is illustrated on some real datasets.

## 1 Introduction

In recent times, mixture models with skew component distributions have received increasing attention. They provide a powerful tool for the modelling and analysis of heterogeneous data with distributions that exhibit non-normal features. These models adopt component densities that can take flexible distributional shapes such as asymmetry and heavy-tailedness. Some notable contributions include mixture modelling with component densities that belong, for example, to the family of skew elliptical distributions (Wang et al., 2009, Lin, 2010, Frühwirth-Schnatter and Pyne, 2010, Cabral et al., 2012, Lin et al., 2014, Lee and McLachlan, 2013c, 2014, 2016a), to the family of generalized hyperbolic distributions (Karlis and Santourian, 2009), and also the multiple-scaled version of some of these distributions (Forbes and Wraith, 2014, Wraith and Forbes, 2015). Among these, the skew normal (SN) and skew $t$ (ST) mixture models are enjoying increasing popularity, with many fruitful applications in a range of important fields such as biology, finance, imaging, medicine, pharmacy, and social sciences (Abanto-Valle et al., 2015, Asparouhov and Muthén, 2015, Bernardi, 2013, Hu et al., 2013, Lee and McLachlan, 2013a,b, 2016b, Lee et al., 2014, 2016b, Lin et al., 2015a,b, Muthén and Asparouhov, 2014, Pyne et al., 2009, 2014, 2015, Riggi and Ingrassia, 2013, Schaarschmidt et al., 2015).

Maximum likelihood (ML) estimation for the parameters of the finite mixture of SN and ST distributions can be carried out via the Expectation-Maximization (EM) algorithm (see the aforementioned references). However, the E-step for such models involves the calculation of the moments of the truncated normal or $t$-distribution. Although these quantities can be expressed in terms of the cumulative distribution function (cdf) of the normal or $t$-distribution, the latter are multidimensional integrals that are computationally expensive to evaluate. The time required for current routines to evaluate these integrals increases as the dimension of the integral increases. This can lead to slow performance for high-dimensional and/or very large datasets.

To reduce the computation time for fitting these models, Lee et al. (2016a) presented a simple multithreaded version of the EM algorithm that spread the computation of the E- and M-steps across $g$ threads, where $g$ is the number of components in a finite mixture model. Their approach was focussed on simplicity and ease of implementation, requiring minimal changes to existing coding. However, further improvement in time performance can be achieved by allowing for the splitting of the data into blocks, whereby a larger number of threads can be run in parallel.

In this paper, we present a block version of the EM algorithm for the fitting of multivariate SN and ST mixture models. Due to the structure of the EM algorithm for these mixture models, conditional expectations on the E-step can be performed independently for each observation in the data and for each component of the mixture model. In a similar manner, the expressions for the updates of the parameters on the M-step can also be computed independently for each individual component. Furthermore, with slight modification, the computations in the M-step can also be split up across different blocks of the data. Thus one may schedule these blocks to be executed concurrently on different threads with an additional step to combine results obtained from the threads at the end of an EM iteration. This approach allows existing implementation to be easily scaled up to support the analysis of large datasets and better utilize resources from machines with multiple cores or processors.

For illustrative purposes, we adopt the canonical fundamental characterization of the skew normal and skew $t$-distributions as component densities of our mixture model. These are referred to as the canonical fundamental skew normal (CFUSN) and canonical fundamental skew $t$ (CFUST) distributions. They represent a fairly general characterization that encompasses some of the more commonly used characterizations of the SN and ST distributions, including the classical formulation by Azzalini and Capitanio (2003) and the version by Sahu et al. (2003). For further details and discussions on the CFUST distribution and its link to various other versions of the multivariate ST distribution, the reader is referred to the papers by Lee and McLachlan (2016a) and McLachlan and Lee (2014, 2016). An EM algorithm for the fitting of finite mixtures of CFUSN and CFUST distributions was presented in the thesis Leemaqz (2014) and in the recent paper by Lee and McLachlan (2016a), respectively. The EM algorithm in the latter paper was implemented in an R package EMMIXcskew (Lee and McLachlan, 2015b) with details presented in Lee and McLachlan (2015a). Our proposed approach for block and parallel implementation will be applied to these two versions of the EM algorithm.

The rest of this paper is organised as follows. In Section 2, we briefly describe the CFUSN and CFUST distributions and their nested models. Section 3 provides an overview of the EM algorithm for fitting mixtures of CFUSN and CFUST distributions. In Section 4, we present the details of a block implementation of this algorithm. Its effectiveness will be demonstrated on some real datasets in Section 5. A summary is then given in Section

6.

## 2 Finite mixtures of skew normal and skew $t$-distributions

Skew normal and skew $t$ distributions are generalizations of the normal and $t$-distributions, respectively. They have extra parameters for the regulation of the skewness of the distribution. Various characterizations of the SN and ST distributions can be defined depending on the mechanism used to introduce skewness to the $t$-distribution. The canonical fundamental skew normal distribution (CFUSN) distribution was introduced as a member of the family of fundamental skew distribution by Arellano-Valle and Genton (2005). This is a fairly general characterization of the skew distribution that encompasses many other existing formulations of skew distributions.

The density of a CFUSN distribution can be expressed in terms of the product of a (multivariate) normal density and the distribution function of another normal distribution. More formally, let $\boldsymbol{Y}$ be a $p$-dimensional random vector that follows the CFUSN distribution. Then its density can be expressed as

$$f_{\text{CFUSN}}(\boldsymbol{y};\boldsymbol{\mu},\boldsymbol{\Sigma},\boldsymbol{\Delta}) = 2^q \, \phi_p(\boldsymbol{y};\boldsymbol{\mu},\boldsymbol{\Omega}) \, \Phi_q(\boldsymbol{c}(\boldsymbol{y});\boldsymbol{0},\boldsymbol{\Lambda}), \tag{1}$$

where

$$\begin{aligned}
\boldsymbol{\Omega} &= \boldsymbol{\Sigma} + \boldsymbol{\Delta}\boldsymbol{\Delta}^\top, \\
\boldsymbol{c}(\boldsymbol{y}) &= \boldsymbol{\Delta}^\top \boldsymbol{\Omega}^{-1}(\boldsymbol{y}-\boldsymbol{\mu}), \\
\boldsymbol{\Lambda} &= \boldsymbol{I}_q - \boldsymbol{\Delta}^\top \boldsymbol{\Omega}^{-1}\boldsymbol{\Delta}, \\
d(\boldsymbol{y}) &= (\boldsymbol{y}-\boldsymbol{\mu})^\top \boldsymbol{\Omega}^{-1}(\boldsymbol{y}-\boldsymbol{\mu}).
\end{aligned}$$

In the above, $\phi_p(\cdot;\boldsymbol{\mu},\boldsymbol{\Omega})$ denotes the density of a $p$-dimensional normal distribution with mean vector $\boldsymbol{\mu}$ and covariance matrix $\boldsymbol{\Omega}$, and $\Phi_p(\cdot;\boldsymbol{\mu},\boldsymbol{\Omega})$ denotes its corresponding cumulative distribution function (cdf). It can observed from (1) that the CFUSN distribution has parameters $\boldsymbol{\mu}$, $\boldsymbol{\Sigma}$, and $\boldsymbol{\Delta}$. The $p$-dimensional vector $\boldsymbol{\mu}$ are location parameters, $\boldsymbol{\Sigma}$ is a positive definite scale matrix, and $\boldsymbol{\Delta}$ is a $p \times q$ matrix of skewness parameters. Note that $q$ is not necessarily smaller than or equal to $p$, although it is typically not taken to be larger than $p$ in practice.

In a similar way, the canonical fundamental skew $t$ (CFUST) distribution can be expressed as a product of a multivariate $t$-density and the cdf of a $t$-distribution. Its density is given by

$$f_{\text{CFUST}}(\boldsymbol{y};\boldsymbol{\mu},\boldsymbol{\Sigma},\boldsymbol{\Delta},\nu) = 2^q \, t_p(\boldsymbol{y};\boldsymbol{\mu},\boldsymbol{\Omega},\nu) \, T_q\left(\boldsymbol{c}(\boldsymbol{y})\sqrt{\frac{\nu+p}{\nu+d(\boldsymbol{y})}};\boldsymbol{0},\boldsymbol{\Lambda},\nu+p\right) \tag{2}$$

where $t_p(\boldsymbol{y};\boldsymbol{\mu},\boldsymbol{\Omega},\nu)$ denotes the $p$-dimensional $t$-distribution with location parameter $\boldsymbol{\mu}$, scale matrix $\boldsymbol{\Omega}$, and degrees of freedom $\nu$, and $T_p(.;\boldsymbol{\mu},\boldsymbol{\Omega},\nu)$ denotes its corresponding cdf. Compared to the CFUSN distribution, the CFUST distribution has an additional scalar parameter $\nu$ which regulates the tails of the distribution. Note that the CFUSN distribution is a limiting case of the CFUST distribution, as $\nu \to \infty$.

As mentioned previously, the CFUSN and CFUST distributions are fairly flexible in shape and include some commonly used distributions as special and/or limiting cases. This includes the normal, Cauchy, and $t$-distributions which can be obtained from (2) by taking $\boldsymbol{\Delta} = \mathbf{0}$ and letting $\nu \to \infty$ (for the normal distribution) or $\nu = 1$ (for the Cauchy distribution). Moreover, by imposing certain constraints on $\boldsymbol{\Delta}$ in the CFUSN and CFUST densities, we can obtain other characterizations of the SN and ST distributions such as those given by Azzalini and Capitanio (2003), Branco and Dey (2001), Gupta (2003), Lachos et al. (2010), Pyne et al. (2009), Sahu et al. (2003); see Lee and McLachlan (2013c) for further details.

Finite mixture models provide a convenient method to account for unobserved heterogeneity in the data. It is defined as a convex combination of component distributions. This provides a natural representation of the heterogeneity in the data, where each component of the mixture model is usually taken to correspond to the distribution of one of the subpopulations in the data. The density of a $g$-component finite mixture model takes the form

$$f(\boldsymbol{y}; \boldsymbol{\Psi}) = \sum_{h=1}^{g} \pi_h f_h(\boldsymbol{y}; \boldsymbol{\theta}_h), \qquad (3)$$

where $\pi_h$ ($h = 1, \ldots, g$) are the mixing proportions and $f_h(\cdot)$ denotes the density of the $h$th mixture component of the mixture model. The mixing proportions are non-negative and sum to one; that is, they satisfy $\pi_h \geq 0$ and $\sum_{h=1}^{g} \pi_h = 1$. The vector $\boldsymbol{\Psi} = (\pi_1, \ldots, \pi_{g-1}, \boldsymbol{\theta}_1^T, \ldots, \boldsymbol{\theta}_g^T)$ contains all the unknown parameters of the mixture model, with $\boldsymbol{\theta}_h$ containing the unknown parameters of the $h$th component. Common choices for the component density $f_h(\cdot)$ are the normal and $t$-distributions. In this paper, we shall adopt the more general CFUSN or CFUST distribution as the component density of our mixture model. These two models are referred to here as the finite mixture of CFUSN distributions (FM-CFUSN) and finite mixture of CFUST distributions (FM-CFUST), respectively.

## 3 Fitting skew normal and skew $t$-mixture models via the EM algorithm

As in the case of normal and $t$-mixture models, the parameters of the FM-CFUSN and FM-CFUST models can be estimated by maximum likelihood (ML) via the Expectation-Maximization (EM) algorithm. The technical details of the EM algorithm for the case of the FM-CFUSN and FM-CFUST models can be found in the recent work of Leemaqz (2014) and Lee and McLachlan (2016a), respectively. For clarity, we briefly describe the work flow of the EM algorithm here. The algorithm begins with an initialization step which produces as output initial (crude) estimates of the parameters of the model and other relevant information such as the (initial) partition of the data and the (initial) log likelihood value. It then enters an iterative loop that consists of alternating between the E- and M-steps. The output of the E-step is used as input of the M-step, and vice versa. At the end of each EM-iteration, the current results are checked against a stopping criterion which terminates the loop if the criterion is satisfied and otherwise loops back to the E-step. On exiting the loop, the results of the latest M-step are updated. A summary of these steps is given below.

1. **Initialization step**: Obtain initial estimates of the parameters, an initial partition of the data, and the initial log likelihood value.

2. **E-step**: Obtain estimates of the conditional expectations based on the current estimates of the parameters.

3. **M-step**: Compute estimates of the parameters based on the output of the E-step.

4. **Stopping criterion**: Check that the stopping criterion is satisfied. If so, return the output of the M-step; otherwise go to the E-step.

We now give further details of these steps for the FM-CFUSN and FM-CFUST models.

## 3.1 E-step

To establish notation, let $\boldsymbol{\Psi}^{(k)}$ denote the current estimate of $\boldsymbol{\Psi}$ after the $k$th iteration of the EM algorithm. Also, let the subscript $h$ denote the $h$th component of the mixture model; for example, $\boldsymbol{\mu}_h$ is the location vector for the $h$th component density. Hence, the parameters $\boldsymbol{\mu}_h^{(k)}$, $\boldsymbol{\Omega}_h^{(k)}$, and $\boldsymbol{\Delta}_h^{(k)}$ denote the current estimates of the corresponding parameters of the $h$th component after completion of the E- and M-steps on the $k$th iteration. We let also $\boldsymbol{y}_j$ denote the $j$th observation ($j = 1, \ldots, n$).

It can be shown that on the $(k+1)$th iteration, the E-step for the FM-CFUSN model requires the following three conditional expectations to be calculated,

$$\tau_{hj}^{(k)} = \frac{\pi_h^{(k)} f_{\text{CFUSN}}(\boldsymbol{y}_j; \boldsymbol{\mu}_h^{(k)}, \boldsymbol{\Sigma}_h^{(k)}, \boldsymbol{\Delta}_h^{(k)})}{\sum_{h=1}^g f_{\text{CFUSN}}(\boldsymbol{y}_j; \boldsymbol{\mu}_h^{(k)}, \boldsymbol{\Sigma}_h^{(k)}, \boldsymbol{\Delta}_h^{(k)})}, \qquad (4)$$

$$\boldsymbol{e}_{1hj}^{(k)} = E(\boldsymbol{X}_{hj} \mid \boldsymbol{y}_j), \qquad (5)$$

$$\boldsymbol{e}_{2hj}^{(k)} = E(\boldsymbol{X}_{hj}\boldsymbol{X}_{hj} \mid \boldsymbol{y}_j), \qquad (6)$$

where $f_{\text{CFUSN}}$ denotes the density of a CFUSN distribution as defined in (1), and $\boldsymbol{X}_{hj}$ follows the truncated multivariate normal distribution with mean vector $\boldsymbol{q}_{hj}^{(k)}$ and covariance matrix $\boldsymbol{\Lambda}_h^{(k)}$, truncated to the positive hyperspace; that is,

$$\boldsymbol{X}_{hj} \mid \boldsymbol{y}_j \sim tN_q\left(\boldsymbol{c}_{hj}^{(k)}, \boldsymbol{\Lambda}_h^{(k)}; \mathbb{R}^+\right).$$

It can be observed from (5) and (6) above that these two conditional expectations are the first and second moments of $\boldsymbol{X}_{hj}$. Formulae for these two moments were given by Tallis (1961) expressed in terms of the normal cdf.

In the case of the FM-CFUST model, there are five conditional expectations on the E-step, which are given by

$$\tau_{hj}^{(k)} = \frac{\pi_h^{(k)} f_{\text{CFUST}}(\boldsymbol{y}_j; \boldsymbol{\mu}_h^{(k)}, \boldsymbol{\Sigma}_h^{(k)}, \boldsymbol{\Delta}_h^{(k)})}{\sum_{h=1}^g f_{\text{CFUST}}(\boldsymbol{y}_j; \boldsymbol{\mu}_h^{(k)}, \boldsymbol{\Sigma}_h^{(k)}, \boldsymbol{\Delta}_h^{(k)})}, \qquad (7)$$

$$w_{hj}^{(k)} = \left(\frac{\nu_h^{(k)} + p}{\nu_h^{(k)} + d_h^{(k)}(\boldsymbol{y}_j)}\right) \frac{T_q\left(\boldsymbol{c}_{hj}^{(k)}\sqrt{\frac{\nu_h^{(k)}+p+2}{\nu_h^{(k)}d_h^{(k)}(\boldsymbol{y}_j)}}; \boldsymbol{0}, \boldsymbol{\Lambda}_h^{(k)}, \nu_h^{(k)}+p+2\right)}{T_q\left(\boldsymbol{c}_{hj}^{(k)}\sqrt{\frac{\nu_h^{(k)}+p}{\nu_h^{(k)}d_h^{(k)}(\boldsymbol{y}_j)}}; \boldsymbol{0}, \boldsymbol{\Lambda}_h^{(k)}, \nu_h^{(k)}+p\right)}, \qquad (8)$$

$$e_{1hj}^{(k)} = w_{hj}^{(k)} - \log\left(\frac{\nu_h^{(k)} + d_h^{(k)}(\boldsymbol{y}_j)}{2}\right) - \left(\frac{\nu_h^{(k)} + p}{\nu_h^{(k)} + d_h^{(k)}(\boldsymbol{y}_j)}\right) + \psi\left(\frac{\nu_h^{(k)} + p}{2}\right), \qquad (9)$$

$$\boldsymbol{e}_{2,hj}^{(k)} = w_{hj}^{(k)} E_{\boldsymbol{\Psi}^{(k)}}[\boldsymbol{u}_{hj} \mid \boldsymbol{y}_j], \qquad (10)$$

$$\boldsymbol{e}_{3hj}^{(k)} = w_{hj}^{(k)} E_{\boldsymbol{\Psi}^{(k)}}[\boldsymbol{u}_{hj}\boldsymbol{u}_{hj}^T \mid \boldsymbol{y}_j], \qquad (11)$$

where $\boldsymbol{U}_{hj}$ given $\boldsymbol{y}_j$ has a $q$-dimensional truncated $t$-distribution given by

$$\boldsymbol{U}_{hj} \mid \boldsymbol{y}_j \ \sim \ tt_q\left(\boldsymbol{c}_{hj}^{(k)}, \left(\frac{\nu_h^{(k)} + d_h(\boldsymbol{y}_j)}{\nu_h^{(k)} + p + 2}\right)\boldsymbol{\Lambda}_h^{(k)}, \nu_h^{(k)} + p + 2; \mathbb{R}^+\right).$$

Similar to the case of the FM-CFUSN model, it can be observed from (10) and (11) that these two conditional expectations are the first and second moments of $\boldsymbol{U}_{hj}$. They can be evaluated using a similar approach to that for the truncated normal distribution, which allows them to be expressed in terms of the multivariate $t$-cdf. The explicit expressions are detailed in Lee and McLachlan (2015a).

It is useful to note here that these conditional expectations are evaluated separately for each $h = 1, \ldots, g$ and $j = 1, \ldots, n$, and that the order in which they are evaluated (over $h$ and $j$) is not important. In particular, in the case of the FM-CFUSN model, the quantities (4) to (6) can be evaluated independently and in no particular order as they do not depend on one another. However, in the case of the FM-CFUST model, evaluation of (9) to (11) requires (8). Hence $e_{1hj}^{(k)}$, $\boldsymbol{e}_{2hj}^{(k)}$, and $\boldsymbol{e}_{3hj}^{(k)}$ must be computed after $w_{hj}^{(k)}$, although they can computed in any order after $w_{hj}^{(k)}$ is obtained.

## 3.2 M-step

On the $(k+1)$th iteration of the the M-step, the current estimate of $\boldsymbol{\Psi}$, $\boldsymbol{\Psi}^{(k)}$, is updated to $\boldsymbol{\Psi}^{(k+1)}$, which is chosen to globally maximize the so-called $Q$-function over $\boldsymbol{\Psi}$. For the FM-CFUSN model, this leads to updates of the parameters $\boldsymbol{\mu}_h$, $\boldsymbol{\Sigma}_h$, and $\boldsymbol{\Delta}_h$, which are given, respectively, by

$$\pi_h^{(k+1)} \ = \ \frac{1}{n}\sum_{j=1}^n \tau_{hj}^{(k)}, \tag{12}$$

$$\boldsymbol{\mu}_h^{(k+1)} \ = \ \frac{\sum_{j=1}^n \tau_{hj}^{(k)}\boldsymbol{y}_j - \boldsymbol{\Delta}_h^{(k)}\sum_{j=1}^n \tau_{hj}^{(k)}\boldsymbol{e}_{1hj}^{(k)}}{\sum_{j=1}^n \tau_{hj}^{(k)}}, \tag{13}$$

$$\boldsymbol{\Delta}_h^{(k+1)} \ = \ \left[\sum_{j=1}^n \tau_{hj}^{(k)}\left(\boldsymbol{y}_j - \boldsymbol{\mu}_h^{(k+1)}\right)\boldsymbol{e}_{1hj}^{(k)T}\right]\left[\sum_{j=1}^n \tau_{hj}^{(k)}\boldsymbol{e}_{2hj}^{(k)}\right]^{-1}, \tag{14}$$

$$\boldsymbol{\Sigma}_h^{(k+1)} \ = \ \left[\sum_{j=1}^n \tau_{hj}^{(k)}\left(\boldsymbol{y}_j - \boldsymbol{\mu}_h^{(k+1)} - \boldsymbol{\Delta}_h^{(k)}\boldsymbol{e}_{1hj}^{(k)}\right)\left(\boldsymbol{y}_j - \boldsymbol{\mu}_h^{(k+1)} - \boldsymbol{\Delta}_h^{(k)}\boldsymbol{e}_{1hj}^{(k)}\right)^T\right.$$

$$\left. + \boldsymbol{\Delta}_h^{(k)}\sum_{j=1}^n \tau_{hj}^{(k)}\left(\boldsymbol{e}_{2hj}^{(k)} - \boldsymbol{e}_{1hj}^{(k)}\boldsymbol{e}_{1hj}^{(k)T}\right)\boldsymbol{\Delta}_h^{(k)T}\right]\left(\sum_{j=1}^n \tau_{hj}^{(k)}\right)^{-1}. \tag{15}$$

As can be observed, the M-step for the FM-CFUSN is given in closed form.

For the FM-CFUST model, the M-step computes updates of $\pi_h$, $\boldsymbol{\mu}_h$, $\boldsymbol{\Sigma}_h$, $\boldsymbol{\Delta}_h$, and $\nu_h$. With the exception of $\nu_h$, the expressions for these parameters are given in closed form, as follows,

$$\pi_h^{(k+1)} = \frac{1}{n}\sum_{j=1}^{n}\tau_{hj}^{(k)}, \qquad (16)$$

$$\boldsymbol{\mu}_h^{(k+1)} = \frac{\sum_{j=1}^{n}\tau_{hj}w_{hj}^{(k)}\boldsymbol{y}_j - \boldsymbol{\Delta}_h^{(k)}\sum_{j=1}^{n}\tau_{hj}^{(k)}\boldsymbol{e}_{2hj}^{(k)}}{\sum_{j=1^n}\tau_{hj}^{(k)}w_{hj}^{(k)}}, \qquad (17)$$

$$\boldsymbol{\Delta}_h^{(k+1)} = \left[\sum_{j=1}^{n}\tau_{hj}^{(k)}\left(\boldsymbol{y}_j - \boldsymbol{\mu}_h^{(k+1)}\right)\boldsymbol{e}_{2hj}^{(k)\top}\right]\left[\sum_{j=1}^{n}\tau_{hj}^{(k)}\boldsymbol{e}_{3hj}^{(k)}\right]^{-1}, \qquad (18)$$

$$\boldsymbol{\Sigma}_h^{(k+1)} = \left[\sum_{j=1}^{n}\tau_{hj}^{(k)}\right]^{-1}\left\{\sum_{j=1}^{n}\tau_{hj}^{(k)}\left[w_{hj}^{(k)}\left(\boldsymbol{y}_j - \boldsymbol{\mu}_h^{(k+1)}\right)\left(\boldsymbol{y}_j - \boldsymbol{\mu}_h^{(k+1)}\right)^T\right.\right.$$
$$\left.\left. - \boldsymbol{\Delta}_h^{(k+1)}\boldsymbol{e}_{3hj}^{(k)\top}\boldsymbol{\Delta}_h^{(k+1)\top}\right]\right\}. \qquad (19)$$

An update $\nu_h^{(k+1)}$ of the degrees of freedom $\nu_h$ is obtained by solving the following equation,

$$0 = \left(\sum_{h=1}^{n}\tau_{hj}^{(k)}\right)\left[\log\left(\frac{\nu_h^{(k+1)}}{2}\right) - \psi\left(\frac{\nu_h^{(k+1)}}{2}\right) + 1\right] - \sum_{j=1}^{n}\tau_{hj}^{(k)}\left(e_{1hj}^{(k)} - w_{hj}^{(k)}\right), \qquad (20)$$

where $\psi(\cdot)$ denotes the digamma function.

Concerning the order in which the updates of the parameters is to evaluated, it can be observed from the expressions given above that the computation of the estimates of $\boldsymbol{\mu}_h$, $\boldsymbol{\Sigma}_h$, and $\boldsymbol{\Delta}_h$ depend on each another. The usual approach to undertake this is to adopt the ECM extension of the EM algorithm, which allows these parameters to be evaluated individually conditional on the other parameters being fixed at their current estimated values. This implies we can evaluate these parameters in any order.

## 4 A block implementation of the EM algorithm

As mentioned previously, the structure of the EM algorithm for mixture models allows for the independent computation of most of the expressions in the E- and M-steps. This implies the conditional expectations can be evaluated for each observation separately and so in parallel on different threads.

Suppose now that the data are partitioned into $N$ blocks ($N \leq n$) of approximately the same size (that is, having a similar number of observations in each block). Let $B_b$ denote the set of indices for observations in the $b$th block ($b = 1, \ldots, N$). The block EM algorithm allows the computation of the E-step and partial M-step to be processed on $N$ concurrent threads. Note that these calculations are the most computationally intensive part of the EM algorithm for the FM-CFUSN and FM-CFUST models. The remaining calculations on the M-step are then performed on a single thread, which involves only calculations that are not computationally expensive. A summary of this work flow is shown in Figure 1 and details of each step are described below.

## 4.1 The E-step for the $b$th block

Due to the structure of the EM algorithm for the fitting of finite mixture models, all of the conditional expectations on the E-step can be performed independently for each observation. As can be observed from (4) to (6) for the case of the FM-CFUSN model, and from (7) to (11) for the case of the FM-CFUST model, these conditional expectations for an observation $\boldsymbol{y}_j$ do not involve the other observations. This suggests that $\tau_{hj}^{(k)}$, $\boldsymbol{e}_{1,hj}^{(k)}$, and $\boldsymbol{e}_{2,hj}^{(k)}$ (or $\tau_{hj}^{(k)}$, $w_{hj}^{(k)}$, $e_{1,hj}^{(k)}$, $\boldsymbol{e}_{2,hj}^{(k)}$, and $\boldsymbol{e}_{3,hj}^{(k)}$ in the case of the FM-CFUST model) can be performed simultaneously on $N$ different threads, where a thread $b$ ($b = 1, \ldots, N$) is responsible for the calculation of those expectations with indices $j \in B_b$. Note that these sets of computations are independent and do not require communication between the threads.

One may observe from (4) and (7) that $\tau_{hj}^{(k)}$ involves the evaluation of the density function for observation $\boldsymbol{y}_j$, denoted by $f_{\text{FM-CFUSN}}(\boldsymbol{y}_j; \boldsymbol{\Psi}^{(k)})$ and $f_{\text{FM-CFUST}}(\boldsymbol{y}_j; \boldsymbol{\Psi}^{(k)})$ respectively for the FM-CFUSN and FM-CFUST models. Note that this is the density value of the mixture model and not the individual components. This quantity is used also in the calculation of the likelihood function at the end of an EM iteration. To avoid re-evaluation of $f_{\text{FM-CFUSN}}(\boldsymbol{y}_j; \boldsymbol{\Psi}^{(k)})$ and $f_{\text{FM-CFUST}}(\boldsymbol{y}_j; \boldsymbol{\Psi}^{(k)})$ after the M-step, the threads are requested to return these quantities together with the other output of the E-step.

In summary, it follows that the input for thread $b$ consists of a partition of the data and the current estimates of the parameters of the model. At the end of the process, we collect from thread $b$ the values of the conditional expectations as listed in Section 3.1 as well as the summation of the denominator of $\tau_{hj}^{(k)}$.

This leads to the following set of tasks for thread $b$:

1. Compute the conditional expectations as listed in Section 3.1 for $j \in B_b$ and $h = 1, \ldots, g$. For the FM-CFUSN model, this includes (4), (5), and (6). For the FM-CFUST model, this includes (7) to (11).

2. Compute the sum of the denominators of the $\tau_{hj}^{(k)}$ across $j \in B_b$, that is,

$$L_b^{(k)} = \sum_{j \in B_b} f_{\text{FM-CFUSN}}(\boldsymbol{y}_j; \boldsymbol{\Psi}^{(k)})$$

and

$$L_b^{(k)} = \sum_{j \in B_b} f_{\text{FM-CFUST}}(\boldsymbol{y}_j; \boldsymbol{\Psi}^{(k)})$$

for the FM-CFUSN and FM-CFUST models, respectively.

## 4.2 The M1-step for the $b$th block

From Section 3.2, it can be observed that the expressions on the M-step involve the summation of various terms containing the $\boldsymbol{y}_j$ and the conditional expectations obtained from the E-step. To speed up the computation, these summations can be computed firstly across the $N$ threads, where thread $b$ ($b = 1, \ldots, N$) is requested to compute the summation for $j \in B_b$. The results will be combined later in the M2-step. For the FM-CFUSN model, it follows that the expressions in the M-step involve six different summations of the conditional expectations obtained form the E-step. We denote these summations by $m_{1h}$ to $\boldsymbol{m}_{6h}$. The following set of quantities is produced in thread $b$ ($b = 1, \ldots, N$):

$$m_{1h}^{(k)} = \sum_{j \in B_b} \tau_{hj}^{(k)}, \tag{21}$$

$$\boldsymbol{m}_{2h}^{(k)} = \sum_{j \in B_b} \tau_{hj}^{(k)} \boldsymbol{e}_{1,hj}^{(k)}, \tag{22}$$

$$\boldsymbol{m}_{3h}^{(k)} = \sum_{j \in B_b} \tau_{hj}^{(k)} \boldsymbol{e}_{2,hj}^{(k)}, \tag{23}$$

$$\boldsymbol{m}_{4h}^{(k)} = \sum_{j \in B_b} \tau_{hj}^{(k)} \boldsymbol{y}_{hj}, \tag{24}$$

$$\boldsymbol{m}_{5h}^{(k)} = \sum_{j \in B_b} \tau_{hj}^{(k)} \boldsymbol{y}_{hj} \boldsymbol{e}_{1,hj}^{(k)^T}, \tag{25}$$

$$\boldsymbol{m}_{6h}^{(k)} = \sum_{j \in B_b} \tau_{hj}^{(k)} \boldsymbol{y}_{hj} \boldsymbol{y}_{hj}^T. \tag{26}$$

Note that the order in which (21) to (26) are computed is not important.

A similar set of quantities need to be evaluated for the FM-CFUST model. From (17) to (20), they are given by

$$m_{1h}^{(k)} = \sum_{j \in B_b} \tau_{hj}^{(k)}, \tag{27}$$

$$m_{2h}^{(k)} = \sum_{j \in B_b} \tau_{hj}^{(k)} w_{hj}^{(k)} \tag{28}$$

$$m_{3h}^{(k)} = \sum_{j \in B_b} \tau_{hj}^{(k)} e_{2,hj}^{(k)}, \tag{29}$$

$$\boldsymbol{m}_{4h}^{(k)} = \sum_{j \in B_b} \tau_{hj}^{(k)} \boldsymbol{e}_{3,hj}^{(k)}, \tag{30}$$

$$\boldsymbol{m}_{5h}^{(k)} = \sum_{j \in B_b} \tau_{hj}^{(k)} \boldsymbol{e}_{4,hj}^{(k)}, \tag{31}$$

$$\boldsymbol{m}_{6h}^{(k)} = \sum_{j \in B_b} \tau_{hj}^{(k)} w_{hj}^{(k)} \boldsymbol{y}_{hj}, \tag{32}$$

$$\boldsymbol{m}_{7h}^{(k)} = \sum_{j \in B_b} \tau_{hj}^{(k)} w_{hj}^{(k)} \boldsymbol{y}_{hj} \boldsymbol{y}_{hj}^T, \tag{33}$$

$$\boldsymbol{m}_{8h}^{(k)} = \sum_{j \in B_b} \tau_{hj}^{(k)} \boldsymbol{y}_{hj} \boldsymbol{e}_{3,hj}^{(k)^T}. \tag{34}$$

In summary, this leads to the following set of tasks for thread $b$:

1. Compute the summations described above for each component. For the FM-CFUSN model, this includes (21) to (26). For the FM-CFUST model, this includes (27) to (34).

2. Return the summations and $L_{bh}^{(k)}$ to the master thread.

Note that for each partition $b$ of the data, the M1-step is performed immediately after the E-step on the same thread.

## 4.3 The M2-step for the master thread

As the summation of the conditional expectations have already been computed in the M1-step, the remaining work in the M-step is to simply combine these summations to obtain an updated estimates of the parameters of the model. Note that this involves only simple summation and matrix multiplication which should require (almost) negligible time compared to the E-step. Thus the M2-step is performed by the master thread. For the FM-CFUSN model, these updates are given by the following expressions:

$$\pi_h^{(k+1)} = \frac{m_{1h}^{(k)}}{n}, \tag{35}$$

$$\boldsymbol{\mu}_h^{(k+1)} = \frac{\boldsymbol{m}_{4h}^{(k)} - \boldsymbol{\Delta}_h^{(k)} \boldsymbol{m}_{2h}^{(k)}}{m_{1h}^{(k)}}, \tag{36}$$

$$\boldsymbol{\Delta}_h^{(k+1)} = \left(\boldsymbol{m}_{5h}^{(k)} - \boldsymbol{\mu}_h^{(k+1)} \boldsymbol{m}_{2h}^{(k)T}\right) \left(\boldsymbol{m}_{3h}^{(k)}\right)^{-1}, \tag{37}$$

$$\boldsymbol{\Sigma}_h^{(k+1)} = \frac{1}{m_{1h}^{(k)}} \left(\boldsymbol{m}_{6h}^{(k)} - \boldsymbol{m}_{4h}^{(k)} \boldsymbol{\mu}_h^{(k)T} - \boldsymbol{\mu}_h^{(k+1)} \boldsymbol{m}_{4h}^{(k)T} + m_{1h}^{(k)} \boldsymbol{\mu}_h \boldsymbol{\mu}_h^T - \boldsymbol{\Delta}_h^{(k+1)} \boldsymbol{m}_{3h}^{(k)} \boldsymbol{\Delta}_h^{(k+1)}\right). \tag{38}$$

In a similar way, the updates of the parameters for the FM-CFUST model are given by:

$$\pi_h^{(k+1)} = \frac{m_{1h}^{(k)}}{n}, \tag{39}$$

$$\boldsymbol{\mu}_h^{(k+1)} = \frac{\boldsymbol{m}_{6h}^{(k)} - \boldsymbol{\Delta}_h^{(k)} \boldsymbol{m}_{4h}^{(k)}}{m_{2h}^{(k)}}, \tag{40}$$

$$\boldsymbol{\Delta}_h^{(k+1)} = \left(\boldsymbol{m}_{8h}^{(k)} - \boldsymbol{\mu}_h^{(k+1)} \boldsymbol{m}_{4h}^{(k)T}\right) \left(\boldsymbol{m}_{5h}^{(k)}\right)^{-1}, \tag{41}$$

$$\boldsymbol{\Sigma}_h^{(k+1)} = \frac{1}{m_{1h}^{(k)}} \left(\boldsymbol{m}_{7h}^{(k)} - \boldsymbol{m}_{6h}^{(k)} \boldsymbol{\mu}_h^{(k)T} - \boldsymbol{\mu}_h^{(k+1)} \boldsymbol{m}_{6h}^{(k)T} - \boldsymbol{\mu}_h^{(k+1)} \boldsymbol{m}_{4h}^{(k)T} + m_{2h}^{(k)} \boldsymbol{\mu}_h \boldsymbol{\mu}_h^T \right.$$
$$\left. - \boldsymbol{\Delta}_h^{(k+1)} \boldsymbol{m}_{5h}^{(k)} \boldsymbol{\Delta}_h^{(k+1)}\right), \tag{42}$$

and the updates of degrees of freedom is obtained by solving

$$0 = \left(\sum_{h=1}^{n} \tau_{hj}^{(k)}\right) \left[\log\left(\frac{\nu_h^{(k+1)}}{2}\right) - \psi\left(\frac{\nu_h^{(k+1)}}{2}\right) + 1\right] - \left(m_{3h}^{(k)} - m_{2h}^{(k)}\right). \tag{43}$$

Thus the M2-step consists of the following tasks for the master thread:

1. Compute the update of the estimates of the parameters defined above for each component. For the FM-CFUSN model, these are given by (35) to (38). For the FM-CFUST model, these are given by (39) to (43).

2. Obtain the current value of likelihood function $L^{(k)}$ by computing the sum of the $L_b^{(k)}$ (see (44) for the FM-CFUSN model or (45) for the FM-CFUST model).

Note that for each partition $b$ of the data, the M1-step is performed immediately after the E-step on the same thread.

## 4.4 The work flow of the block EM algorithm

At the end of each EM iteration, a check for convergence needs to be performed to determine whether the algorithm should be stopped. To proceed, the current value of the likelihood function needs to be computed. This is given by the sum of the logarithm of the density of the FM-CFUSN or FM-CFUST models evaluated at the data points; that is, it is given by

$$L^{(k)} = \sum_{j=1}^{n} \log f_{\text{FM-CFUSN}}(\boldsymbol{y}_j; \boldsymbol{\Psi}^{(k)}) = \sum_{b=1}^{N} \log L_b^{(k)} \tag{44}$$

and

$$L^{(k)} = \sum_{j=1}^{n} \log f_{\text{FM-CFUST}}(\boldsymbol{y}_j; \boldsymbol{\Psi}^{(k)}) = \sum_{b=1}^{N} \log L_b^{(k)} \tag{45}$$

respectively, for the FM-CFUSN and FM-CFUST models. It can be observed from (44) that the terms in the summation are the same as in the denominator of $\tau_{hj}^{(k)}$. Thus, as mentioned previously, one can save time by not recomputing the density values here. In other words, $L^{(k)}$ is obtained by summing the logarithm of density values given by the output of the $N$ threads. This task can be performed as part of the M2-step (see the above section). However, it should be noted that this gives the value of $L^{(k)}$ rather than $L^{(k+1)}$.

When the stopping criterion is met, one may compute a partial E-step, that is, involving only the computation of $\tau_{hj}^{(k+1)}$ across the $N$ threads, to obtain the final update of $\tau_{hj}$ and the latest estimate of the likelihood function. Hence, if the stopping criterion is met after the $k$th iteration, the following tasks are performed:

1. For thread $b$ ($b = 1, \ldots, N$):

    (a) Compute the conditional expectation $\tau_{hj}^{(k+1)}$ for $h = 1, \ldots, g$ and $j \in B_b$.

    (b) Compute $L_b^{(k+1)}$, the sum of the denominators of the $\tau_{hj}^{(k)}$ across $j \in B_b$.

2. For the master thread, compute $L^{(k+1)}$ by summing up the $L_b^{(k+1)}$.

## 5 Applications

To demonstrate the use of the proposed method, we consider the two real datasets used in Lee et al. (2016a). For both datasets, we take the number of components $g$ to be known. Here we are interested in the performance gain of the block implementation described in Section 4. Hence we will compare the reduction in computation time against the traditional (serial) implementation.

### 5.1 Iris dataset

The first dataset is the well-known Iris dataset (Fisher, 1936) which consists of 150 observations and four variables. As there are three species of Iris in the data, we took $g = 3$ when fitting the mixture models. One important issue with parallel implementations of algorithms such as the EM algorithm is the choice of the number of blocks ($N$). Since there are overheads involved in setting up the parallel process, there is a limit in the performance gain of the parallel implementation. This limit will depend on the machine/system used and the data.

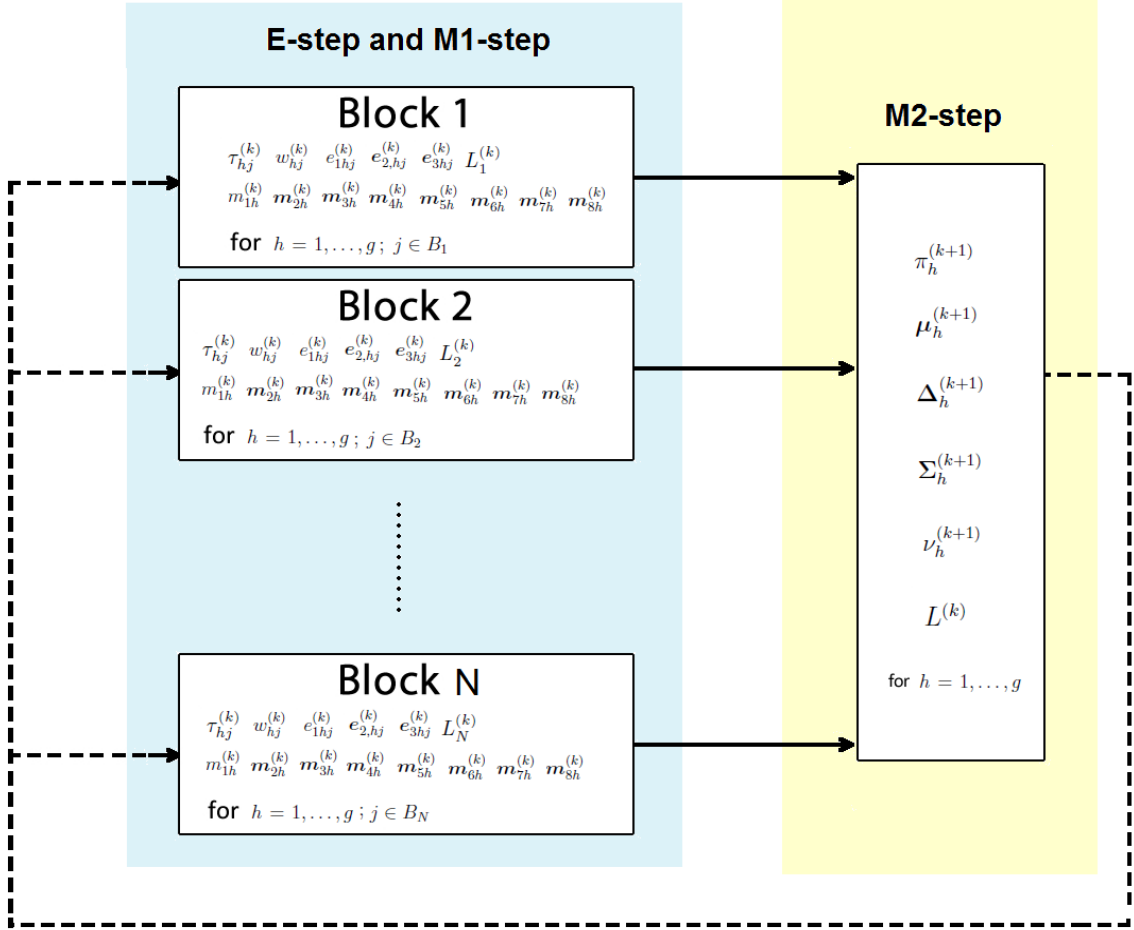

Figure 1: Workflow of the basic block EM algorithm for the FM-CFUST model. Each block in the first column is responsible for the calculation of the E-step and the first partial M-step (M1-step) for a portion of the observations. The vertical block in the second column collects results for the M1-step from all blocks and evaluates second partial M-step (M2-step) for all components. This includes the calculation of the likelihood value as well other metrics specified by the user.

| Number of blocks | Reduction in time (%) | |
| :---: | :---: | :---: |
| (B) | FM-CFUSN | FM-CFUST |
| 2 | 24.81 | 3.90 |
| 3 | 35.70 | 10.15 |
| 4 | 40.67 | 42.71 |
| 5 | 46.63 | 54.08 |
| 6 | 52.57 | 59.32 |
| 7 | 49.61 | 63.00 |
| 8 | 50.62 | 66.46 |
| 9 | 53.59 | 64.57 |
| 10 | 46.65 | 63.67 |

Table 1: Reduction in time (%) of the FM-CFUSN and FM-CFUST models for different number of blocks on the Iris dataset.

Here we experiment with $N$ ranging from 2 to 10 on a standard quad-core PC. Figure 2 shows percentage reduction in time relative to the serial implementation (bottom panel). Table 1 shows the percentage reduction in time for the block implementation and combined implementation relative to the standard EM implementation for both the FM-CFUSN and FM-CFUST models. It can be observed that the optimal speedup for both the FM-CFUSN and FM-CFUST models is achieved when $N = 8$ in this case. The percentage reduction in computation time is somewhat similar for both models (see Figure 2), although the total computation time is considerably higher for the FM-CFUST model (not shown).

## 5.2 HSCT dataset

We consider also the hematopoietic stem cell transplant (HSCT) dataset used in Lee et al. (2016a), which consists of over 6000 observations in $p = 4$ dimensions. As in Lee et al. (2016a), we applied our algorithm to the HSCT dataset with $g = 4$. Similar to the above experiment, we set $B$ to range from 2 to 10. With this dataset, the optimal speedup appears to be at $N = 9$ blocks for the FM-CFUST model (see Figure 3 and Table 2). For the FM-CFUSN model, the best performance in time was achieved at $N = 8$ blocks in this experiment. In addition, it can be observed that the performance gain is higher for this dataset compared to the Iris dataset, as the latter is a smaller dataset.

A further remark on Figure 3 is that there seems to be little if anything to be gained in the performance time for $N$ beyond 6 or so. The trend starts to decrease at $N = 10$, suggesting the overheads involved are starting to have a significant impact on the total computation time. The results suggest that having $N = 8$ blocks of the data would provide a good balance between excessive overheads and optimal performance under the setup in this experiment.

It is of interest to note that Lee et al. (2016a) reported a reduction in time of around 60% using their multithreaded implementation with four concurrent threads. As can be observed from Table 2, a greater reduction in computation time can be achieved with three threads using the block implementation.

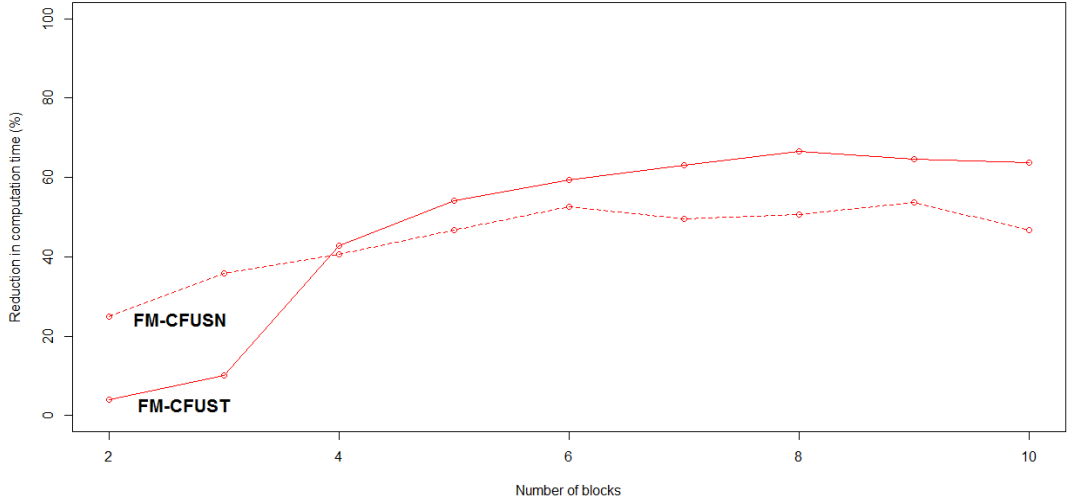

Figure 2: Performance gain on the Iris dataset using the FM-CFUSN (dash line) and (solid line) FM-CFUST models. The reduction in time (in percentage) relative to the serial implementation is shown for $N = 2, \ldots, 10$ blocks. The first point in the plot (at $N=1$) represents the time obtained using the traditional (serial) implementation.

| Number of blocks | Reduction in time (%) | |
| (B) | FM-CFUSN | FM-CFUST |
| --- | --- | --- |
| 2 | 62.87 | 51.24 |
| 3 | 73.12 | 67.50 |
| 4 | 76.77 | 74.58 |
| 5 | 78.05 | 79.50 |
| 6 | 77.53 | 82.35 |
| 7 | 80.02 | 85.27 |
| 8 | 80.25 | 86.86 |
| 9 | 79.79 | 87.48 |
| 10 | 79.27 | 86.80 |

Table 2: Reduction in time (%) of the FM-CFUSN and FM-CFUST models for different number of blocks on the HSCT dataset.

# 6  Conclusions

We have presented a new parallelization scheme for implementing the EM algorithm for the fitting of finite mixtures of CFUSN and CFUST distributions. Our block implementation of the EM algorithm allows the computation load of the E-step and first part (M1) of the M-step of the EM algorithm to be spread across $N$ concurrent threads. The performance gain of our approach was demonstrated on two real datasets. In the smaller Iris dataset, the reduction in computation time ranged from around 4% to 66% when $N$ (the number of blocks) was varied from 2 to 10, with the optimal speedup observed at $N = 8$. For the larger HSCT dataset, the reduction in computation time was between approximately 50% to 87%, and the best performance gain was observed when $N = 9$. In both cases, a greater performance gain was achieved compared to the recently proposed multithreaded implementation. Another advantage of our block implementation is that it allows the user to specify the number of threads $N$ to be used, whereas in the former implementation $N$ is fixed to be equal to the number of components $g$ of the mixture model. This provides greater flexibility to the user and allows better utilisation of processing resources available from the machine.

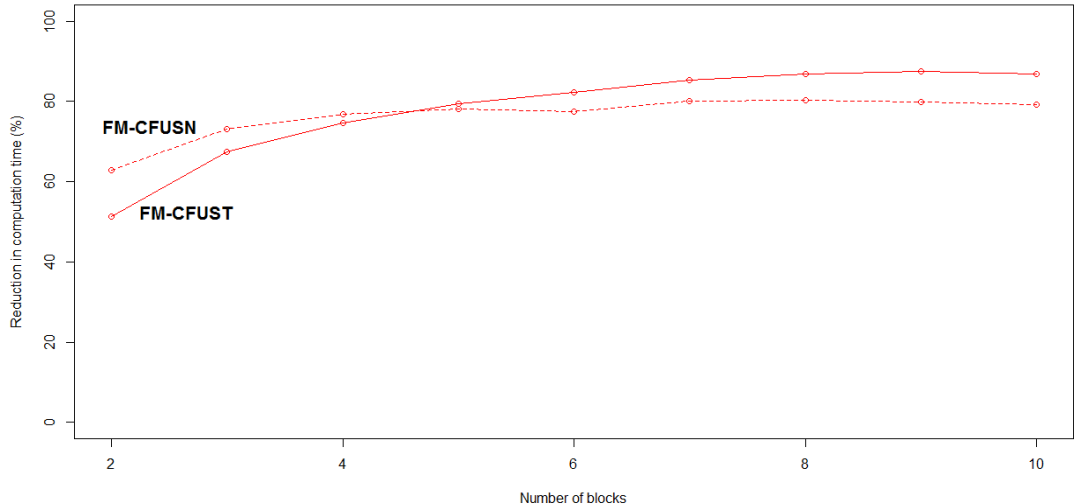

Figure 3: Performance gain on the HSCT dataset using the FM-CFUSN (dash line) and FM-CFUST (solid line) models. The reduction in time (in percentage) relative to the serial implementation is shown for $N = 2, \ldots, 10$ blocks. The first point in the plot (at $N=1$) represents the time obtained using the traditional (serial) implementation.